\documentclass[10pt]{article}

\usepackage{latexsym}
\usepackage{graphics}
\begin{document}
\begin{center}
{\Large \bf Energy bands and Wannier-Mott excitons \\
in Zn(P$_{1-x}$As$_{x}$)$_{2}$ and Zn$_{1-x}$Cd$_{x}$P$_{2}$ crystals}

\vspace{0.5cm}

O.A. Yeshchenko, M.M. Biliy, Z.Z. Yanchuk

\vspace{0.5cm}

{\em Physics Department, Taras Shevchenko National Kyiv University \\
6 Akademik Glushkov prosp., 03127 Kyiv, Ukraine \\
E-mail: yes@mail.univ.kiev.ua}

\end{center}

\begin{abstract}
Excitonic absorption, reflection and photoluminescence spectra of mixed Zn(P$_{1-x}$As$_{x}$)$_{2}$
crystals over the full range of $x$ ($0 \leq x \leq 1$) and Zn$_{1-x}$Cd$_{x}$P$_{2}$ crystals at 
$0 \leq x \leq 0.05$ have been studied at low temperatures (1.8 K). The decrease of the energy 
gap in Zn(P$_{1-x}$As$_{x}$)$_{2}$ at the increase of $x$ occurs slightly sublinearly. The rydbergs
of excitonic series in this crystals decrease as well, and the dependences $Ry(x)$ for all
series are strongly superlinear at small $x$. In Zn$_{1-x}$Cd$_{x}$P$_{2}$ crystals the energy gap 
and rydbergs decrease at the increase of $x$ (at $0 \leq x \leq 0.05$) as well. The dependences 
of $E_{g}$ and $Ry$ on $x$ are considerably stronger in Zn(P$_{1-x}$As$_{x}$)$_{2}$ than in 
Zn$_{1-x}$Cd$_{x}$P$_{2}$. At the increase of $x$ the half-width of excitonic absorption lines 
increases monotonically in both type crystals that is evidence of the increasing role of 
fluctuations of crystal potential. 
\end{abstract}

\section{Introduction}
\label{intro}

In Ref. \cite{PhysB} we performed the spectroscopic studies of an influence of substitution 
of phosphorus by arsenic on the structure of valence and conduction  bands, and on the parameters of Wannier-Mott excitons in mixed crystals (solid 
solutions) of isovalent substitution Zn(P$_{1-x}$As$_{x}$)$_{2}$ at low levels of substitution 
$x \leq 0.05$. In present work we have proceeded with these studies at $0 \leq x \leq 1$. By the other 
hand we have performed here the similar study (in comparison to Zn(P$_{1-x}$As$_{x}$)$_{2}$) of
another type mixed crystals of isovalent substitution Zn$_{1-x}$Cd$_{x}$P$_{2}$ at small 
substitution levels $x \leq 0.05$. As we know, the Zn$_{1-x}$Cd$_{x}$P$_{2}$ crystals were not studied
earlier. Both Zn(P$_{1-x}$As$_{x}$)$_{2}$ and Zn$_{1-x}$Cd$_{x}$P$_{2}$ belong to the mixed crystals of
$A^{II}B^{V}$ type, which are poorly investigated.

$\beta$-ZnP$_{2}$ (further ZnP$_{2}$) and ZnAs$_{2}$ crystals are strong\-ly anisotropic direct-gap 
semiconductors (energy gap: 1.6026 eV for ZnP$_{2}$ and 1.052 eV for ZnAs$_{2}$), which are characterized 
by the same symmetry group $C_{2h}^{5}$ (monoclinic syngony). 
Besides the symmetry of the lattice, the similarity of ZnP$_{2}$ and ZnAs$_{2}$ exists in the structure of 
energy bands and exciton states, namely, three excitonic series are observed in the absorption spectra of 
these crystals: dipole allowed C-series at ${\bf E} \parallel Z({\bf c})$ polarization originating from $S$-states of 
C-exciton (this series is observed in reflection spectra as well), forbidden B-series at
${\bf E} \perp Z({\bf c})$ polarization originating from $S$-states of B-exciton, and partially allowed A-series 
at ${\bf E} \parallel X$ polarization originating from $S$-states of A-exciton (see e.g. 
Refs. \cite{PSS,Pevtsov,Dopovidi} for ZnP$_{2}$ and Refs. \cite{Mudr,ZnAs2UFZh,Moroz} for ZnAs$_{2}$). 
In the photoluminescence (PL) spectra of these crystals at ${\bf E} \parallel Z({\bf c})$ polarization, 
a series of lines caused by the radiative transitions from the ground and excited states of allowed 
C-exciton is observed (see e.g. Ref. \cite{PSSLiq} for ZnP$_{2}$ and Ref. \cite{Mudr} for 
ZnAs$_{2}$). Besides this emission series, the so-called B-line is observed in the PL spectra of ZnP$_{2}$ 
at ${\bf E} \perp Z({\bf c})$. B-line occurs due to the radiative transitions from the ground state of forbidden 
B-exciton and corresponds to B$_{1}$-line of absorption B-series. 

In contrast to ZnP$_{2}$ and ZnAs$_{2}$ crystals, CdP$_{2}$ is indi\-rect-gap semiconductor (energy gap: 2.155 eV: see, e.g. \cite{CdP2}),
which is characterized by the different symmetry of lattice: symmetry group is $D_{4}^{4}$ for right-rotating 
and $D_{4}^{8}$ for left-rotating modification (tetragonal syngony). Therefore, the comparison study
of an influence of substitution of $P$ by $As$ in Zn(P$_{1-x}$As$_{x}$)$_{2}$ and of $Zn$ by $Cd$ in Zn$_{1-x}$Cd$_{x}$P$_{2}$
on parameters of energy bands and excitonic states seems to be rather interesting.

The technological operations of growing of Zn(P$_{1-x}$As$_{x}$)$_{2}$ and Zn$_{1-x}$Cd$_{x}$P$_{2}$ crystals were carried out 
according to described in Ref. \cite{PhysB}.

\section{Zn(P$_{1-x}$As$_{x}$)$_{2}$ crystals over the full range of $x$}
\label{sec1}

\begin{figure*}[p]
\resizebox{1\textwidth}{!}{%
  \includegraphics{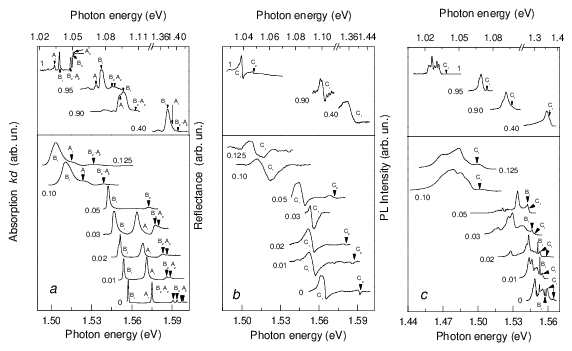}
}
\caption{Optical spectra of Zn(P$_{1-x}$As$_{x}$)$_{2}$ crystals at temperature 1.8 K. (a) Absorption spectra. Observation conditions: ${\bf q} \perp (110)$, 
${\bf E} \perp Z({\bf c})$ - for crystals with any $x$ except $x = 0.05$; ${\bf q} \perp (100)$, ${\bf E} \perp Z({\bf c})$ - for crystals with $x = 0.05$. 
(b) Reflection spectra. Observation conditions: ${\bf q} \perp (100)$, ${\bf E} \parallel Z({\bf c})$.
(c) Photoluminescence spectra. Observation conditions: ${\bf q} \perp (100)$.}
\label{fig1}       
\end{figure*}

In the present work the low-temperature (1.8 K) absorption, reflection and photoluminescence spectra of 
Zn(P$_{1-x}$As$_{x}$)$_{2}$ crystals have been studied at the following levels of substitution of $P$ by $As$: 
$x = $ 0.01, 0.02, 0.03, 0.05, 0.10, 0.125, 0.40, 0.90, 0.95. Respective spectra as well as spectra
of pure ZnP$_{2}$ ($x = 0$) and ZnAs$_{2}$ ($x = 1$) are presented in Fig. \ref{fig1}. As one can expect,
the Zn(P$_{1-x}$As$_{x}$)$_{2}$ crystals are direct-gap semiconductors as well as ZnP$_{2}$ and ZnAs$_{2}$.
One can see from the figure, that in the mixed crystals the same excitonic C-, B- and A-series are observed, 
as in pure crystals. Let us note the doublet structure of an absorption $n = 1$ line of B-series in crystals 
with $x = 0.02$. Proceeding from intensities and half-widths of the components of this doublet, we have made 
a conclusion, that narrow high-energy component is $n = 1$ line of B-series. An origin of low-energy component, 
which is missing for crystals with $x \not= 0.02$, is not clear. One can see from absorption and reflection
spectra of Zn(P$_{1-x}$As$_{x}$)$_{2}$ crystals that at the increase of $x$ (or at the increase of $1-x$ if to go
from the ZnAs$_{2}$ side)\footnote{Further, writing "the increase of $x$" we mean the increase of $x$ at $x \leq 
0.50$ and the increase of $1-x$ at $x > 0.50$.} the lines corresponding to exciton states with higher $n$ disappear, 
and the lines with $n = 1,2$ broaden. Probably, this fact is a result of "blurring" of the  band edges, which takes place owing 
to fluctuations of crystal potential, caused by chaotic distribution of $As$ ($P$) atoms in sites of lattice at 
substitution of $P$ ($As$) atoms. In PL spectra the emission lines of free excitons can be easily separated from the lines
of localized excitons (including, most likely, emission of excitons autholocalized on fluctuations of crystal 
potential) only at rather low substitution levels. At the increase
of $x$ the emission lines of both free and localized excitons broaden that is due to the
fluctuations of crystal potential and corresponding fluctuations of the parameters of energy bands. It takes an attention 
on itself the fact of considerable increase of intensity of PL spectra at the increase
of $x$. Most probably, this effect is due to two following causes. First one consists in the partial suppression of the 
spatial migration of excitons (due to localization of excitons on fluctuations of crystal potential) and the respective 
decrease of efficiency of non-radiative decay of excitons. Second one consists in the fact that the processes of scattering 
of excitons on the lattice defects favour to more effective radiative decay of excitons.

\begin{figure*}[p]
\resizebox{1\textwidth}{!}{%
  \includegraphics{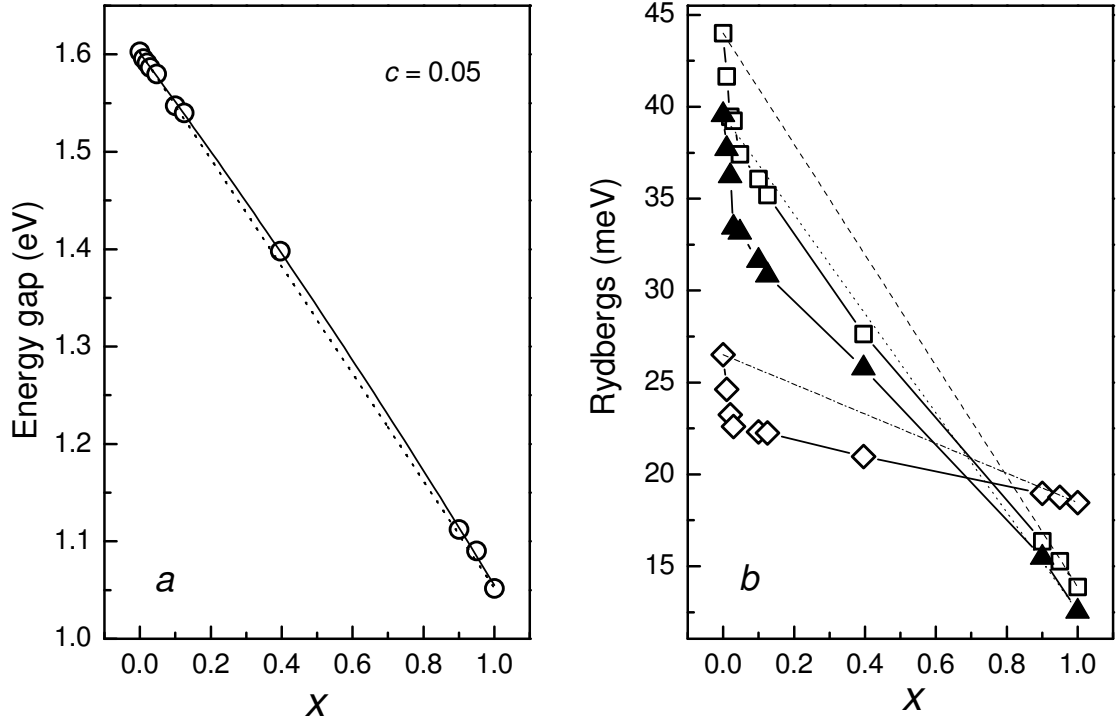}
}
\caption{(a) Dependence of energy gap on $x$ for Zn(P$_{1-x}$As$_{x}$)$_{2}$ crystals. Solid line represents fitting of the experimental points by expression 
(\ref{eq1}), dotted one is the line connecting points of two extreme cases: ZnP$_{2}$ ($ x = 0$) and ZnAs$_{2}$ ($ x = 1$).
(b) Dependences of excitonic rydbergs on $x$ for Zn(P$_{1-x}$As$_{x}$)$_{2}$. Connected open squares - experimental dependence for B-series, connected solid up triangles - the same for C-series, 
connected open diamonds - the same for A-series. Dashed line connects two extreme cases (mentioned above in this caption) for B-series, dotted one - for C-series,
dashed-dotted one - for A-series.}
\label{fig2}       
\end{figure*}

With increase in concentration $x$, spectral lines shift to the low-energy side, which is caused by the decrease of energy gap. 
This shift could be easily expected, taking into account the fact, that in ZnAs$_{2}$ energy gap is 0.55 eV smaller than in 
ZnP$_{2}$. The respective dependence $E_{g}(x)$ is given in Fig. \ref{fig2}(a). The dependence is slightly nonlinear (sublinear).
It has been fitted by the well-known dependence (see, e.g. Ref. \cite{Pikhtin})
\begin{equation}
E_{g}(x) = E_{g1}-(E_{g1}-E_{g2})x+cx(1-x) \, ,
\label{eq1}
\end{equation}
where $E_{g1}$ and $E_{g2}$ are the energy gaps of ZnP$_{2}$ and ZnAs$_{2}$ respectively, and $c$ is the coefficient of 
nonlinearity. The coefficient $c$ was obtained to be 0.05. But besides the trivial decrease of $E_{g}$ with the increase of $x$, 
there is also decrease of the excitonic series rydbergs (see Fig. \ref{fig2}(b)). Values of $E_{g}$ and rydbergs were obtained 
from fitting of excitonic series by simple hydrogenlike dependence: $E(n) = E_{g} - Ry/n^{2}$. One can see from the figure that
the dependences of excitonic rydbergs on $x$ are rather remarkable. These dependences are strongly superlinear at small $x$ 
(close to ZnP$_{2}$) and most linear at $x \rightarrow 1$ (close to ZnAs$_{2}$). Strongest superlinearity takes place at
$x \leq 0.05$. Let us note that the rydbergs of B- and A-series decrease considerably: at crossing from ZnP$_{2}$ to ZnAs$_{2}$
the rydbergs decrease more than in 3 times. Meanwhile, the rydberg of A-series decreases sufficiently less: it decreases in 1.4 times.
Let us note that in ZnP$_{2}$ the $n=1$ state of A-exciton is energy highest, and the $n=1$ state of B-exciton is energy lowest. But, 
due to the slower decrease of $Ry_{A}(x)$ than ones observed for B- and A-series, at substitution levels $x \sim 0.70$ the 
$n=1$ state of A-exciton leaves off to be the energy highest. And in extreme case of ZnAs$_{2}$ an opposite situation takes place:
$A_{1}$-state is energy lowest. The $n=1$ state of allowed C-exciton is energy highest in ZnAs$_{2}$ crystal. There are exist two 
possible causes of such a fast decrease of excitonic rydbergs at small $x$. First one is the decrease of exciton reduced mass at the
increase of $x$. This our assumption is based on the data of the exciton reduced masses in ZnP$_{2}$ ($\mu_{a}=0.45m_{0}$, 
$\mu_{c}=0.10m_{0}$ and $\mu_{b}=0.56m_{0}$ \cite{Engbring}) and ZnAs$_{2}$ ($\mu_{\perp bc}=0.30m_{0}$ \cite{Mudr,Moroz}) as well as on our 
previous evaluation of the dependences of $m_{e}(x)$ and $m_{h}(x)$ at small $x$ \cite{PhysB}. Here, $\mu_{a,b,c}$ is the components of
the reduced mass of exciton in ZnP$_{2}$ in different crystallographic directions, and $\mu_{\perp bc}$ is the component of $\mu$ in the
direction perpendicular to plane (100) (this direction almost coincides with $a$ axis of crystal). Second cause of the fast 
decrease of rydbergs at small $x$ is the possible fast increase of dielectric constant, as in ZnAs$_{2}$ $\epsilon$ is sufficiently
higher: average value $15$ \cite{Mudr} versus $\epsilon_{a}=9.1$, $\epsilon_{c}=9.3$ and $\epsilon_{b}=10.1$ in ZnP$_{2}$. Slower decrease of
excitonic rydbergs at higher $x$ can be most likely explained by possible slower decrease of the reduced exciton mass and increase of the 
dielectric constant at higher $x$. 

\begin{figure*}[p]
\resizebox{0.5\textwidth}{!}{%
  \includegraphics{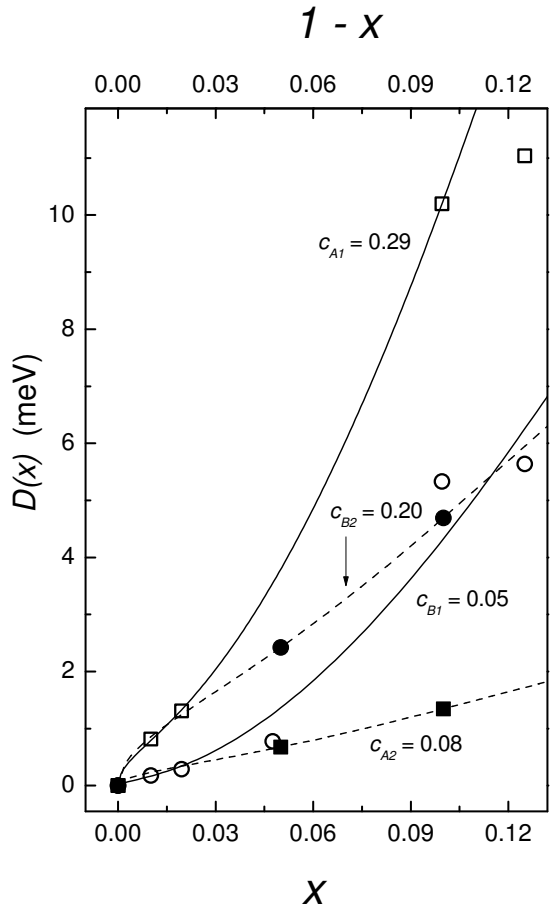}
}
\caption{Dependences of half-widths of excitonic absorption $B_{1}$- and $A_{1}$-lines in Zn(P$_{1-x}$As$_{x}$)$_{2}$ crystals on $x$. Open circles and squares - experimental dependences for $B_{1}$-
and $A_{1}$-lines in crystals close to ZnP$_{2}$ ($x \leq 0.125$: bottom x-axis); solid circles and squares - the same for $B_{1}$- and $A_{1}$-lines in crystals close to ZnAs$_{2}$ ($x \geq 0.90$:
top x-axis). Solid lines represent the fitting of experimental points by expression (\ref{eq4}) for crystals with $x \leq 0.125$; dashed lines - the same for crystals with $x \geq 0.90$.}
\label{fig3}      
\end{figure*}

There were also studied dependences on $x$ of the half-widths of absorption $n=1$ lines of B- and A-series. These dependences are given in 
Fig. \ref{fig3}.\footnote{An absence of data on the half-widths of lines for crystal with $x = 0.03$ is due to the fact that the respective 
sample was rather thick, therefore, full absorption occurred in $B_{1}$- and $A_{1}$-lines. So, it was not possible to determine correctly 
the half-widths of these lines. As we have pointed out above, half-width of $B_{1}$-line for crystal with  $x=0.02$ was determined for the 
high-energy component of doublet.} One can seen, that half-widths of $B_{1}$- and $A_{1}$-lines increase monotonously with the increase of 
$x$. As known, the increase of half-width of exciton lines is the result of fluctuations of crystal potential and respective fluctuations 
of energy gap. The theory of influence of fluctuations of composition $x$ on half-width of exciton absorption lines was developed in 
Ref. \cite{Ablyazov}, where two extreme cases were considered. First one takes place, if the effective size of area of the crystal potential 
fluctuation $R_{D} = \hbar / (2MD)^{1/2}$, where $M$ is the total mass of exciton and $D(x) = W(x) - W(0)$ ($W(x)$ is the half-width of exciton 
line), is much larger than exciton Bohr radius: $R_{D} >> a_{ex}$. Such situation, as a rule, takes place, if the effective masses of electron 
and hole are small and differ slightly: $m_{e} \sim m_{h}$. In this case $D$ should depend on $x$ as
\begin{equation}
D(x) = 0.08\frac{\alpha^{4}M^{3}x^{2}(1-x)^{2}}{\hbar ^{6}N^{2}} \, ,
\label{eq2}
\end{equation}
where $\alpha = dE_{g}/dx$, $N$ is the concentration of sites of lattice, where the substituting atoms can "sit". Such a situation occurs in many 
semiconductors, in particular in $A^{III}B^{V}$ crystals. The second case takes place at $R_{D} << a_{ex}$. It takes place at $m_{h} >> m_{e}$. 
In this case, dependence $D(x)$ has such character
\begin{equation}
D(x) = 0.5\alpha \left( \frac{x(1-x)}{Na_{ex}^{3}} \right)^{1/2} \, .
\label{eq3}
\end{equation}
The experimental dependences $D(x)$ for absorption $B_{1}$- and $A_{1}$-lines of Zn(P$_{1-x}$As$_{x}$)$_{2}$ crystals are presented in Fig. \ref{fig3}.
Likely to data presented in our work \cite{PhysB} the experimental points are badly fitted both by function (\ref{eq2}) and (\ref{eq3}). In ZnP$_{2}$
and ZnAs$_{2}$ crystals the effective size of area of crystal potential fluctuation is about some tens of angstrom, and the Bohr radii of B- and A-excitons
in these crystals are the following: $a_{B} = 16 \AA$, $a_{A} = 29.5 \AA$ in ZnP$_{2}$, and $a_{B} = 34 \AA$, $a_{A} = 25.7 \AA$ in ZnAs$_{2}$, i.e. 
$R_{D} \sim a_{ex}$. Therefore, the extreme conditions $R_{D} >> a_{ex}$ and $R_{D} << a_{ex}$ are not fulfilled, and the intermediate case takes place 
which is, nevertheless, more close to case of Eq. (\ref{eq2}). Therefore, since the intermediate case takes place, the experimental dependences can be 
fitted by the function
\begin{equation}
D(x) = (1-c)D_{1}(x) + cD_{2}(x) \, ,
\label{eq4}
\end{equation}
which is the superposition of function $D_{1}(x)$ of type (\ref{eq2}) and $D_{2}(x)$ of type (\ref{eq3}), $c$ is the weighting factor. One can see from 
Fig. \ref{fig3} that for the mixed crystals close to ZnP$_{2}$, i.e. at $x \rightarrow 0$, the experimental dependence of the half-width of 
$B_{1}$-line on $x$ is fitted by function (\ref{eq2}) rather well, and the contribution of the function (\ref{eq3}) is rather small ($c = 0.05$). The situation is
quite different for the mixed crystals close to ZnAs$_{2}$, i.e. at $x \rightarrow 1$, as the contribution of the function (\ref{eq3}) is considerably
larger ($c = 0.20$). It is rather simple to understand, as the Bohr radius of B-exciton in ZnAs$_{2}$ is about two times larger than in ZnP$_{2}$. So, in ZnP$_{2}$ 
the condition $R_{D} >> a_{ex}$ is fulfilled better than in ZnAs$_{2}$. For the dependence of the half-width of $A_{1}$-lines on $x$ we have the quite opposite
situation. For the mixed crystals close to ZnP$_{2}$, the experimental dependence of the half-width of $A_{1}$-line on $x$ deviates considerably from function 
(\ref{eq2}), and the contribution of the function (\ref{eq3}) is rather large ($c = 0.29$). For the mixed crystals close to ZnAs$_{2}$, the contribution of 
the function (\ref{eq3}) to the fitting function is quite small ($c = 0.08$). The Bohr radius of A-exciton, as we have noted above, almost does not differ in 
ZnP$_{2}$ and ZnAs$_{2}$. So, as the contribution of function (\ref{eq3}) at $x \rightarrow 1$ is smaller than at $x \rightarrow 0$,the condition $R_{D} >> a_{ex}$ 
is fulfilled better for A-exciton in the crystals close to ZnAs$_{2}$. Thus, we can make a conclusion of larger effective size of area of the crystal potential 
fluctuation in crystals close to ZnAs$_{2}$, than in ones close to ZnP$_{2}$.

\section{Zn(P$_{1-x}$As$_{x}$)$_{2}$ and Zn$_{1-x}$Cd$_{x}$P$_{2}$ at small $x$: comparison}
\label{sec:2}

\begin{figure*}[p]
\resizebox{1\textwidth}{!}{%
  \includegraphics{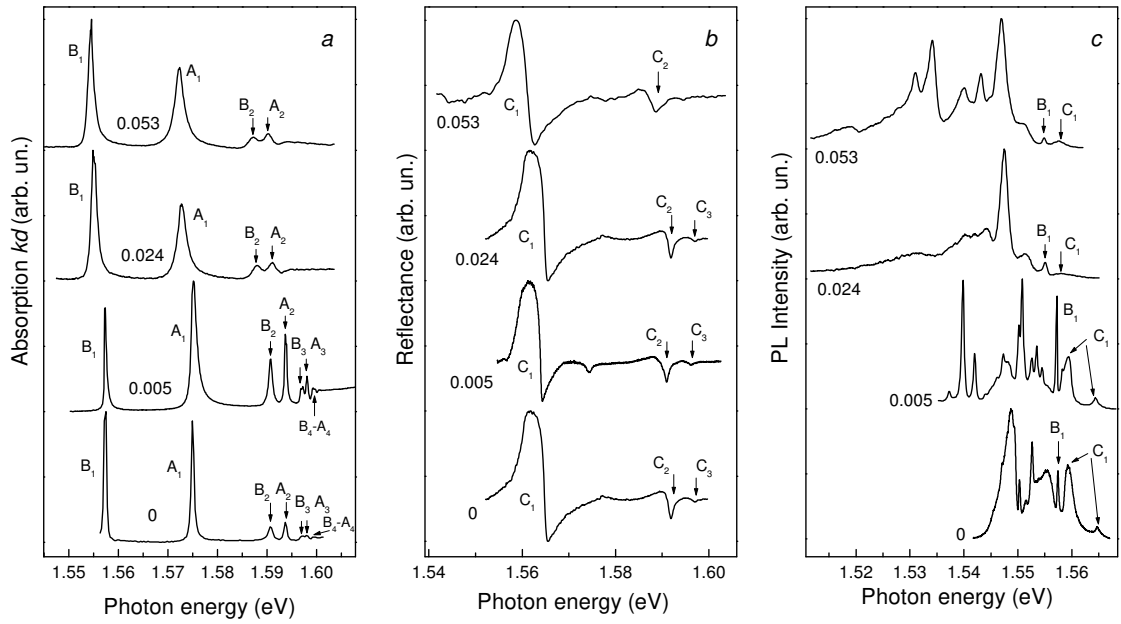}
}
\caption{Optical spectra of Zn$_{1-x}$Cd$_{x}$P$_{2}$ crystals at temperature 1.8 K. (a) Absorption spectra. Observation conditions: ${\bf q} \perp (110)$, 
${\bf E} \perp Z({\bf c})$. 
(b) Reflection spectra. Observation conditions: ${\bf q} \perp (100)$, ${\bf E} \parallel Z({\bf c})$.
(c) Photoluminescence spectra. Observation conditions: ${\bf q} \perp (100)$.}
\label{fig4}       
\end{figure*}

The comparison studies of the low-temperature absorption, reflection and photoluminescence spectra of Zn(P$_{1-x}$As$_{x}$)$_{2}$ and Zn$_{1-x}$Cd$_{x}$P$_{2}$ crystals 
have been performed at small levels of substitution of $Zn$ by $Cd$: $x \leq 0.05$. Respective spectra of Zn(P$_{1-x}$As$_{x}$)$_{2}$ are presented in Fig. \ref{fig1},
and the spectra of Zn$_{1-x}$Cd$_{x}$P$_{2}$ -- in Fig. \ref{fig4}. In spite of the difference of the symmetries of lattice and structure of energy bands of ZnP$_{2}$ 
and CdP$_{2}$ crystals (monoclinic and tetragonal, direct- and indirect-gap respectively: see Sect.~\ref{intro}), at small $x$ Zn$_{1-x}$Cd$_{x}$P$_{2}$ remain 
the direct-gap crystals with monoclinic lattice. One can see from the figure, that in the Zn$_{1-x}$Cd$_{x}$P$_{2}$ crystals the same excitonic C-, B- and A-series 
are observed, as well as in pure ZnP$_{2}$ and mixed Zn(P$_{1-x}$As$_{x}$)$_{2}$crystals. One can see from absorption and reflection spectra that, likely to 
Zn(P$_{1-x}$As$_{x}$)$_{2}$, already at rather small $x$ only the excitonic states with $n = 1,2$ are observed clearly in the spectra of Zn$_{1-x}$Cd$_{x}$P$_{2}$ crystals.
The higher components of exciton spectra disappear, and the lines with $n = 1,2$ broaden. In PL spectra, at the increase of $x$ the emission lines of both free and 
localized excitons broaden. As noted above (see Sect.~\ref{sec1}), all these effects are due to the fluctuations of crystal potential and corresponding fluctuations 
of the parameters of energy bands. Likely to Zn(P$_{1-x}$As$_{x}$)$_{2}$, an effect of considerable increase of intensity of PL spectra at the increase of $x$ takes place
in Zn$_{1-x}$Cd$_{x}$P$_{2}$ crystals too. An explanation of such an effect is given as well in Sect.~\ref{sec1}.

\begin{figure*}[p]
\resizebox{1\textwidth}{!}{%
  \includegraphics{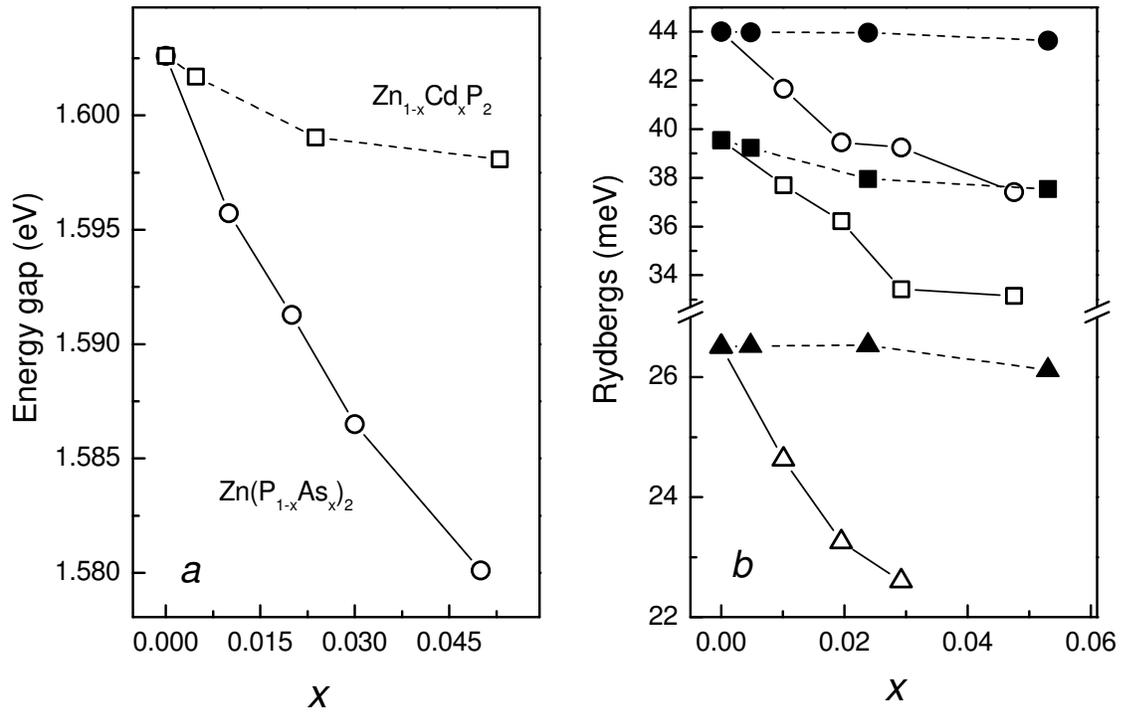}
}
\caption{(a) Dependences of energy gap on $x$ for Zn(P$_{1-x}$As$_{x}$)$_{2}$ (connected open circles) and Zn$_{1-x}$Cd$_{x}$P$_{2}$ crystals (connected open squares).
(b) Dependences of excitonic rydbergs on $x$ for Zn(P$_{1-x}$As$_{x}$)$_{2}$ and Zn$_{1-x}$Cd$_{x}$P$_{2}$. Connected open circles, squares, and up triangles - 
the dependences $Ry(x)$ for B-, C-, and A-series of Zn(P$_{1-x}$As$_{x}$)$_{2}$ crystals respectively; connected solid circles, squares, and up triangles - the dependences 
$Ry(x)$ for B-, C-, and A-series of Zn$_{1-x}$Cd$_{x}$P$_{2}$ crystals respectively.}
\label{fig5}       
\end{figure*}

In spite of the fact that the CdP$_{2}$ crystal has larger energy gap than the ZnP$_{2}$, likely to Zn(P$_{1-x}$As$_{x}$)$_{2}$ there is a decrease of the energy gap 
of Zn$_{1-x}$Cd$_{x}$P$_{2}$ crystals at the increase of $x$ (see Fig. \ref{fig5}(a)). But the decrease of $E_{g}$ in Zn$_{1-x}$Cd$_{x}$P$_{2}$ is much slower. And besides the
similarity in dependences of $E_{g}(x)$, both Zn(P$_{1-x}$As$_{x}$)$_{2}$ and Zn$_{1-x}$Cd$_{x}$P$_{2}$ crystals have the similarity in the dependences of the rydbergs of 
excitons on $x$ as well (see Fig. \ref{fig5}(b)). The rydbergs of all the three A-, B-, and C-series decrease at the increase of $x$. But, also likely to the dependence $E_{g}(x)$, 
the excitonic rydbergs in Zn$_{1-x}$Cd$_{x}$P$_{2}$ decrease considerably slower than in Zn(P$_{1-x}$As$_{x}$)$_{2}$. Considerably weaker dependences of the energy gap and the 
rydbergs on $x$ in Zn$_{1-x}$Cd$_{x}$P$_{2}$ are, most probably, the results of the weaker changes in parameters of the energy bands, in particular in effective masses of electrons 
and holes, and in dielectric constant. But, at higher $x$ one can expect the more interesting and stronger changes in the parameters of energy bands and excitonic states with 
regard to the fact that CdP$_{2}$ and ZnP$_{2}$ have the different symmetry of lattice, and that CdP$_{2}$ is indirect-gap semiconductor. 

\begin{figure*}[p]
\resizebox{0.5\textwidth}{!}{%
  \includegraphics{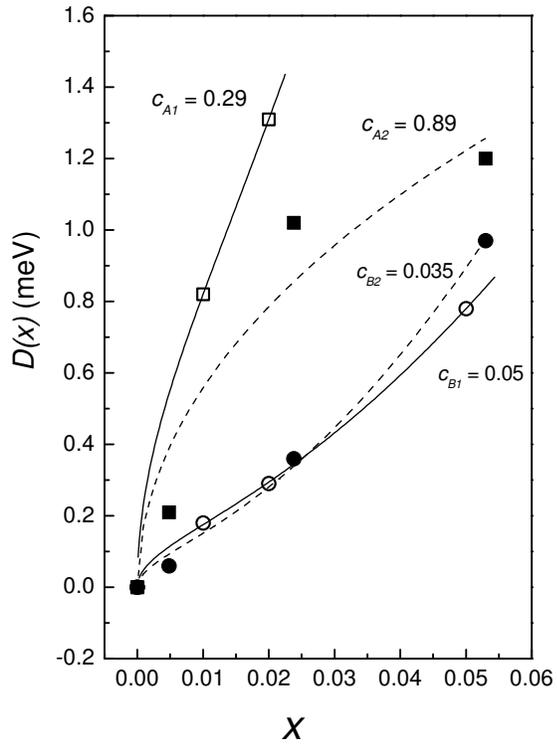}
}
\caption{Dependences of half-widths of excitonic absorption $B_{1}$- and $A_{1}$-lines on $x$: open circles and squares respectively -- in Zn(P$_{1-x}$As$_{x}$)$_{2}$,
and solid circles and squares respectively -- in Zn$_{1-x}$Cd$_{x}$P$_{2}$ crystals. Solid lines represent the fitting of experimental points by expression (\ref{eq4}) 
for Zn(P$_{1-x}$As$_{x}$)$_{2}$ crystals; dashed lines - the same for Zn$_{1-x}$Cd$_{x}$P$_{2}$ crystals.}
\label{fig6}       
\end{figure*}

The dependences of half-widths of absorption $B_{1}$- and $A_{1}$-lines on $x$ have been also studied in Zn$_{1-x}$Cd$_{x}$P$_{2}$ crystals. The results of comparison of 
such dependences with ones observed for Zn(P$_{1-x}$As$_{x}$)$_{2}$ are presented in Fig. \ref{fig6}. One can see that $D(x)$ dependences for both types of the mixed crystals 
are similar for exciton $B_{1}$-line. This dependences are described rather well by the function of Eq. (\ref{eq2}) type, contributions of the function (\ref{eq3}) are
quite small for both type crystals: the weighting factor is $c = 0.05$ for Zn(P$_{1-x}$As$_{x}$)$_{2}$ and $c = 0.035$ for Zn$_{1-x}$Cd$_{x}$P$_{2}$. It is an evidence that in 
both cases the effective size of area of the crystal potential fluctuation $R_{D}$ is rather large comparing to the exciton Bohr radius $a_{ex}$. Likely to 
Zn(P$_{1-x}$As$_{x}$)$_{2}$, in Zn$_{1-x}$Cd$_{x}$P$_{2}$ the dependence of half-width of $A_{1}$-line on $x$ deviates strongly from the function (\ref{eq2}), the contribution
of the function (\ref{eq3}) is large. As we assumed above, it is the result of the fact that A-exciton has considerably lower binding energy comparing to B-exciton, and 
larger Bohr radius correspondingly. So, the condition $R_{D} >> a_{ex}$ is not fulfilled, and the dependence $D(x)$ is not described by the function (\ref{eq2}). At the same
time, there are two remarkable features of $D(x)$ dependence for $A_{1}$-line in Zn$_{1-x}$Cd$_{x}$P$_{2}$. First one is the considerable deviations of 
the experimental points from the fitting function. It is quite remarkable as such deviations are very small for the $D(x)$ dependences for $A_{1}$-line in 
Zn(P$_{1-x}$As$_{x}$)$_{2}$ and for $B_{1}$-line in the mixed crystals of both types. Second one is the fact that the contribution of the function (\ref{eq3}) is determinative
in the fitting function: the weighting factor of the function (\ref{eq3}) is $c = 0.89$. For Zn(P$_{1-x}$As$_{x}$)$_{2}$ the contribution of the function (\ref{eq3}) in 
the fitting function is considerably smaller: $c = 0.29$. At present moment, both above features of $D(x)$ dependence for $A_{1}$-line in Zn$_{1-x}$Cd$_{x}$P$_{2}$ are obscure.
Thus, a comparison of the energy bands and exciton parameters behaviour versus $x$ reveals as similar tendencies and different ones.

\section{Conclusions}
\label{concl}

In conclusion, we have obtained the following results. We have found that, likely to "pure" ZnP$_{2}$ crystal, both in Zn(P$_{1-x}$As$_{x}$)$_{2}$ and Zn$_{1-x}$Cd$_{x}$P$_{2}$ 
mixed crystals the same three excitonic hydrogenlike series (C, B, and A) are observed. At the increase of $x$ in the range of small $x$ ($0 \leq x \leq 0.05$) 
the decrease of the energy gap and exciton rydbergs takes place. The dependences of $E_{g}$ and $Ry$ on $x$ are considerably stronger in Zn(P$_{1-x}$As$_{x}$)$_{2}$ than 
in Zn$_{1-x}$Cd$_{x}$P$_{2}$. 

Besides the comparison study of these crystals at small $x$, the Zn(P$_{1-x}$As$_{x}$)$_{2}$ crystals have been studied over the full range of $x$: $0 \leq x \leq 1$. We have 
obtained the following results for Zn(P$_{1-x}$As$_{x}$)$_{2}$. At the increase of $x$ the energy gap decreases slightly sublinearly. The exciton rydbergs decrease as well. 
The dependences $Ry(x)$ are strongly superlinear at small $x$ (close to ZnP$_{2}$) and most linear at $x \rightarrow 1$ (close to ZnAs$_{2}$). At the crossing from ZnP$_{2}$ 
to ZnAs$_{2}$, the rydbergs of B- and C-series decrease more than in 3 times. Meanwhile, the rydberg of A-series decreases sufficiently less: in 1.4 times. 

At the increase of $x$ the half-width of excitonic absorption lines increases monotonically both in Zn(P$_{1-x}$As$_{x}$)$_{2}$ and Zn$_{1-x}$Cd$_{x}$P$_{2}$ crystals that is 
evidence of the increasing role of fluctuations of crystal potential.


\end{document}